\documentclass[twocolumn]{aastex631}

\suppressAffiliations

\newcommand{\OSU}{\label{OSU} Department of Astronomy, The Ohio State University, 140 West 18th Avenue, Columbus, Ohio 43210, USA}

\newcommand{\Alberta}{\label{Alberta} Department of Physics, University of Alberta, Edmonton, AB T6G 2E1, Canada}

\newcommand{\ANU}{\label{ANU} Research School of Astronomy and Astrophysics, Australian National University, Canberra, ACT 2611, Australia}

\newcommand{\Carnegie}{\label{Carnegi} Observatories of the Carnegie Institution for Science, 813 Santa Barbara Street, Pasadena, CA 91101, USA}

\newcommand{\CITEVA}{\label{CITEVA} Centro de Astronomía (CITEVA), Universidad de Antofagasta, Avenida Angamos 601, Antofagasta, Chile}

\newcommand{\ESO}{\label{ESO} European Southern Observatory, Karl-Schwarzschild Stra{\ss}e 2, D-85748 Garching bei M\"{u}nchen, Germany}

\newcommand{\HD}{\label{HD} Astronomisches Rechen-Institut, Zentrum f\"{u}r Astronomie der Universit\"{a}t Heidelberg, M\"{o}nchhofstra\ss e 12-14, D-69120 Heidelberg, Germany}

\newcommand{\ICRAR}{\label{ICRAR} International Centre for Radio Astronomy Research, University of Western Australia, 35 Stirling Highway, Crawley, WA 6009, Australia}

\newcommand{\IRAM}{\label{IRAM} Institut de Radioastronomie Millim\'{e}trique (IRAM), 300 Rue de la Piscine, F-38406 Saint Martin d'H\`{e}res, France}

\newcommand{\ITA}{\label{ITA} Universit\"{a}t Heidelberg, Zentrum f\"{u}r Astronomie, Institut f\"{u}r Theoretische Astrophysik, Albert-Ueberle-Str 2, D-69120 Heidelberg, Germany}

\newcommand{\IWR}{\label{IWR} Universit\"{a}t Heidelberg, Interdisziplin\"{a}res Zentrum f\"{u}r Wissenschaftliches Rechnen, Im Neuenheimer Feld 205, D-69120 Heidelberg, Germany}

\newcommand{\JHU}{\label{JHU} Department of Physics and Astronomy, The Johns Hopkins University, Baltimore, MD 21218, USA}

\newcommand{\MPE}{\label{MPE} Max-Planck-Institut f\"{u}r extraterrestrische Physik, Giessenbachstra{\ss}e 1, D-85748 Garching, Germany}

\newcommand{\MPIA}{\label{MPIA} Max-Planck-Institut f\"{u}r Astronomie, K\"{o}nigstuhl 17, D-69117, Heidelberg, Germany}

\newcommand{\NRAO}{\label{NRAO} National Radio Astronomy Observatory, 520 Edgemont Road, Charlottesville, VA 22903-2475, USA}

\newcommand{\Oxford}{\label{Oxford} Sub-department of Astrophysics, Department of Physics, University of Oxford, Keble Road, Oxford OX1 3RH, UK}

\newcommand{\UBonn}{\label{UBonn} Argelander-Institut f\"ur Astronomie, Universit\"at Bonn, Auf dem H\"ugel 71, 53121 Bonn, Germany}

\newcommand{\UChile}{\label{UChile} Departamento de Astronom\'{i}a, Universidad de Chile, Camino del Observatorio 1515, Las Condes, Santiago, Chile}

\newcommand{\UGent}{\label{UGent} Sterrenkundig Observatorium, Universiteit Gent, Krijgslaan 281 S9, B-9000 Gent, Belgium}

\newcommand{\ULyon}{\label{ULyon} Univ Lyon, Univ Lyon 1, ENS de Lyon, CNRS, Centre de Recherche Astrophysique de Lyon UMR5574,\\ F-69230 Saint-Genis-Laval, France}

\newcommand{\UWyoming}{\label{UWyoming} Department of Physics and Astronomy, University of Wyoming, Laramie, WY 82071, USA}

\newcommand{\STScIESA}{\label{STScIESA} AURA for the European Space Agency (ESA), Space Telescope Science Institute, 3700 San Martin Drive, Baltimore, MD 21218, USA}

\newcommand{\INAF}{\label{INAF} INAF -- Osservatorio Astrofisico di Arcetri, Largo E. Fermi 5, I-50157, Florence, Italy}

\newcommand{\TKU}{\label{TKU} Department of Physics, Tamkang University, No.151, Yingzhuan Rd., Tamsui Dist., New Taipei City 251301, Taiwan}

\newcommand{\UCSD}{\label{UCSD} Center for Astrophysics \& Space Sciences, Department of Physics, University of California, San Diego, 9500 Gilman Drive, San Diego, CA 92093, USA}

\newcommand{\Ha}{$\mathrm{H\alpha}$}

\newcommand{\SIIHa}{[S~\textsc{ii}]/$\mathrm{H\alpha}$}
\newcommand{\NIIHa}{[N~\textsc{ii}]/$\mathrm{H\alpha}$}
\newcommand{\OIIIHb}{[O~\textsc{iii}]/$\mathrm{H\beta}$}

\newcommand{\SII}{[S~\textsc{ii}]}
\newcommand{\SIII}{[S~\textsc{iii}]}
\newcommand{\SIIISII}{[S~\textsc{iii}]/[S~\textsc{ii}]}
\newcommand{\HII}{H~\textsc{ii}}

\newcommand{\JWST}{\rm {\it JWST}}

\newcommand{\Spitzer}{\rm {\it Spitzer}}

\newcommand{\qPAH}{$q_\mathrm{PAH}$}
\newcommand{\nreg}{1529}

\def\revone{}
\def\revtwo{}

\begin{document}

\title{PHANGS-JWST First Results: Destruction of the PAH molecules in HII regions probed by JWST and MUSE}

\correspondingauthor{Oleg Egorov}\email{oleg.egorov@uni-heidelberg.de}

\author[0000-0002-4755-118X]{Oleg V. Egorov}
\affiliation{\HD}

\author[0000-0001-6551-3091]{Kathryn Kreckel}
\affiliation{\HD}
\author[0000-0002-4378-8534]{Karin M. Sandstrom}
\affiliation{\UCSD}
\author[0000-0002-2545-1700]{Adam~K.~Leroy}
\affiliation{\OSU}
\author[0000-0001-6708-1317]{Simon C.\ O.\ Glover}
 \affiliation{\ITA}
\author[0000-0002-9768-0246]{Brent Groves}
\affiliation{\ICRAR}
\author[0000-0002-8804-0212]{J.~M.~Diederik~Kruijssen}
\affiliation{Cosmic Origins Of Life (COOL) Research DAO, coolresearch.io}
\author[0000-0003-0410-4504]{Ashley.~T.~Barnes}
\affiliation{\UBonn}
\author[0000-0002-2545-5752]{Francesco Belfiore}
\affiliation{\INAF}
\author[0000-0003-0166-9745]{F. Bigiel}
\affiliation{\UBonn}
\author[0000-0003-4218-3944]{Guillermo A. Blanc}
\affiliation{\Carnegie}
\affiliation{\UChile}
\author[0000-0003-0946-6176]{Médéric~Boquien}
\affiliation{\CITEVA}
\author[0000-0001-5301-1326]{Yixian Cao}
\affiliation{\MPE}
\author[0000-0002-5235-5589]{J\'er\'emy Chastenet}
\affiliation{\UGent}
\author[0000-0002-5635-5180]{M\'elanie Chevance}
\affiliation{\ITA}
\affiliation{Cosmic Origins Of Life (COOL) Research DAO, coolresearch.io}
\author[0000-0002-8549-4083]{Enrico Congiu}
\affiliation{\UChile}
\author[0000-0002-5782-9093]{Daniel~A.~Dale}
\affiliation{\UWyoming}
\author[0000-0002-6155-7166]{Eric Emsellem}
\affiliation{\ESO}
\affiliation{\ULyon}
\author[0000-0002-3247-5321]{Kathryn~Grasha}
\affiliation{\ANU}   
\author[0000-0002-0560-3172]{Ralf S.\ Klessen}
 \affiliation{\ITA}
 \affiliation{\IWR}
 \author[0000-0003-3917-6460]{Kirsten L. Larson}
\affiliation{\STScIESA}
 \author[0000-0001-9773-7479]{Daizhong Liu}
\affiliation{\MPE}
\author[0000-0001-7089-7325]{Eric J.\,Murphy}
\affiliation{\NRAO}
\author[0000-0002-1370-6964]{Hsi-An Pan}
\affiliation{\TKU} 
\author[0000-0002-0873-5744]{Ismael Pessa}
\affiliation{\MPIA}
\author[0000-0003-3061-6546]{J\'er\^ome Pety}
\affiliation{\IRAM}
\affiliation{LERMA, Observatoire de Paris, PSL Research University, CNRS, Sorbonne Universit\'es, 75014 Paris}
\author[0000-0002-5204-2259]{Erik Rosolowsky}
\affiliation{\Alberta}
\author[0000-0003-2707-4678]{Fabian Scheuermann}
\affiliation{\HD}
\author[0000-0002-3933-7677]{Eva Schinnerer}
\affiliation{\MPIA}
\author[0000-0002-9183-8102]{Jessica Sutter}
\affiliation{\UCSD}
\affiliation{Leibniz-Institut f\"{u}r Astrophysik Potsdam (AIP), An der Sternwarte 16, 14482 Potsdam, Germany}
\author[0000-0002-8528-7340]{David A. Thilker}
\affiliation{\JHU}
\author[0000-0002-7365-5791]{Elizabeth J. Watkins}
\affiliation{\HD}
\author[0000-0002-0786-7307]{Thomas G. Williams}
\affiliation{\Oxford}
\affiliation{\MPIA}


\begin{abstract}
Polycyclic aromatic hydrocarbons (PAHs) play a critical role in the reprocessing of stellar radiation and in balancing the heating and cooling processes in the interstellar medium (ISM), but appear to be destroyed in \HII\ regions. 
However, the mechanisms driving their destruction are still not completely understood. Using  PHANGS-JWST and PHANGS-MUSE observations,   
we investigate how the PAH fraction changes in about 1500 \HII\ regions across four nearby star-forming galaxies (NGC~628, NGC~1365, NGC~7496, IC~5332). We find a strong anti-correlation between the PAH fraction and the ionization parameter (the ratio between the ionizing photon flux and the hydrogen density)  of \HII\ regions. This relation becomes steeper for more luminous \HII\ regions. The metallicity of \HII\ regions has only a minor impact on these results in our galaxy sample. We find that the PAH fraction decreases with the \Ha\ equivalent width -- a proxy for the age of the \HII\ regions -- although this trend is much weaker than the one identified using the ionization parameter. Our results are consistent with a scenario where hydrogen-ionizing UV radiation is the dominant source of PAH destruction in star-forming regions. 

\end{abstract}

\keywords{Interstellar dust (836) --- H II regions (694) --- Polycyclic aromatic hydrocarbons (1280)	
}


\section{Introduction} \label{sec:intro}

Polycyclic aromatic hydrocarbons (PAHs) are carbon-based macromolecules that are ubiquitous in the interstellar medium (ISM), and are traced by several strong emission features at 3.3, 6.2, 7.7, 8.6, 11.3, 12.7, and 17~$\mu$m \citep{Tielens2008, Li2020}. Together with very small ($20-30$~\AA) dust grains that have been stochastically heated, PAHs can be a dominant contributor to the mid-infrared (mid-IR) spectra of galaxies \citep[e.g.][]{Draine2007}. 
The IR emission from PAHs is produced by the absorption and re-emission of ultraviolet (UV) photons, with PAHs reprocessing as much as 20\% of all stellar UV radiation \citep{Smith2007}. Because of this, they have been proposed as indicators of star formation rate \citep{Calzetti2013}, but their usage for this purpose is complicated by our lack of understanding of the mechanisms for PAH formation and destruction \citep{Whitcomb2020}.

PAHs are formed in the carbon-rich atmospheres of evolved stars \cite[e.g.][]{Latter1991, Cherchneff1992}, though these PAHs alone cannot account for all the PAH emission observed in the ISM \cite[e.g.][]{Matsuura2009, Matsuura2013}. 
The precursor molecules for PAHs have been found in dense clouds \citep{Burkhardt2021}, and high PAH fractions are observed in molecular clouds, which is difficult to explain if they do not form there given cloud lifetimes \citep{Sandstrom2010, Chastenet2019}. 
A popular scenario for PAH formation is 
the shattering of larger dust grains \cite[e.g.][]{Jones1996, Hirashita2009, Seok2014, Wiebe2014}. 
Both shocks and UV radiation play a critical but complex role in the evolution of PAHs. Shocks can increase the PAH fraction by dissociating large grains, but also decrease the PAH abundance by destroying them directly \cite[e.g.][]{OHalloran2006, Micelotta2010}. Similarly, UV radiation excites the IR bands in PAHs, but can also destroy them \citep[e.g.][]{Allain1996, Pavlyuchenkov2013}. 

PAH destruction is thought to be regulated by the radiation field, either by the high intensity of ionizing photons \citep{Montillaud2013} or the hardness\footnote{Various definitions are used in the literature (e.g., the effective temperature, or the spectral index of the UV spectrum). Here we define `hardness' as the ratio of the fluxes over hard and soft UV ranges (see Appendix~\ref{sec:app}).} of the radiation \citep{Madden2006, Gordon2008}. Theoretical works also suggest that PAHs are
subject to sputtering and fragmentation in ionized gas due to
electronic and atomic interactions \citep{Micelotta2010ig, Bocchio2012}.
These processes should destroy PAH molecules, and thus reduce their abundance in \HII\ regions. Observationally, this has been seen as a lack of PAH emission in the interior of Galactic \HII\ regions \cite[e.g.][]{Povich2007}, and an anti-correlation between the PAH fraction and tracers of the hardness and/or intensity of the ionizing radiation on large (global) scales \citep[e.g.][]{Madden2006, Gordon2008, Hunt2010, Lebouteiller2011, Maragkoudakis2018}. 
Hydrogen-ionizing radiation is often considered a natural explanation for the observed decrease of the PAH abundances in low-metallicity galaxies and star-forming complexes \citep[e.g.][]{Engelbracht2005, Madden2006, Engelbracht2008, Khramtsova2013}. 

\begin{figure*}
    \centering
    \includegraphics[width=\linewidth]{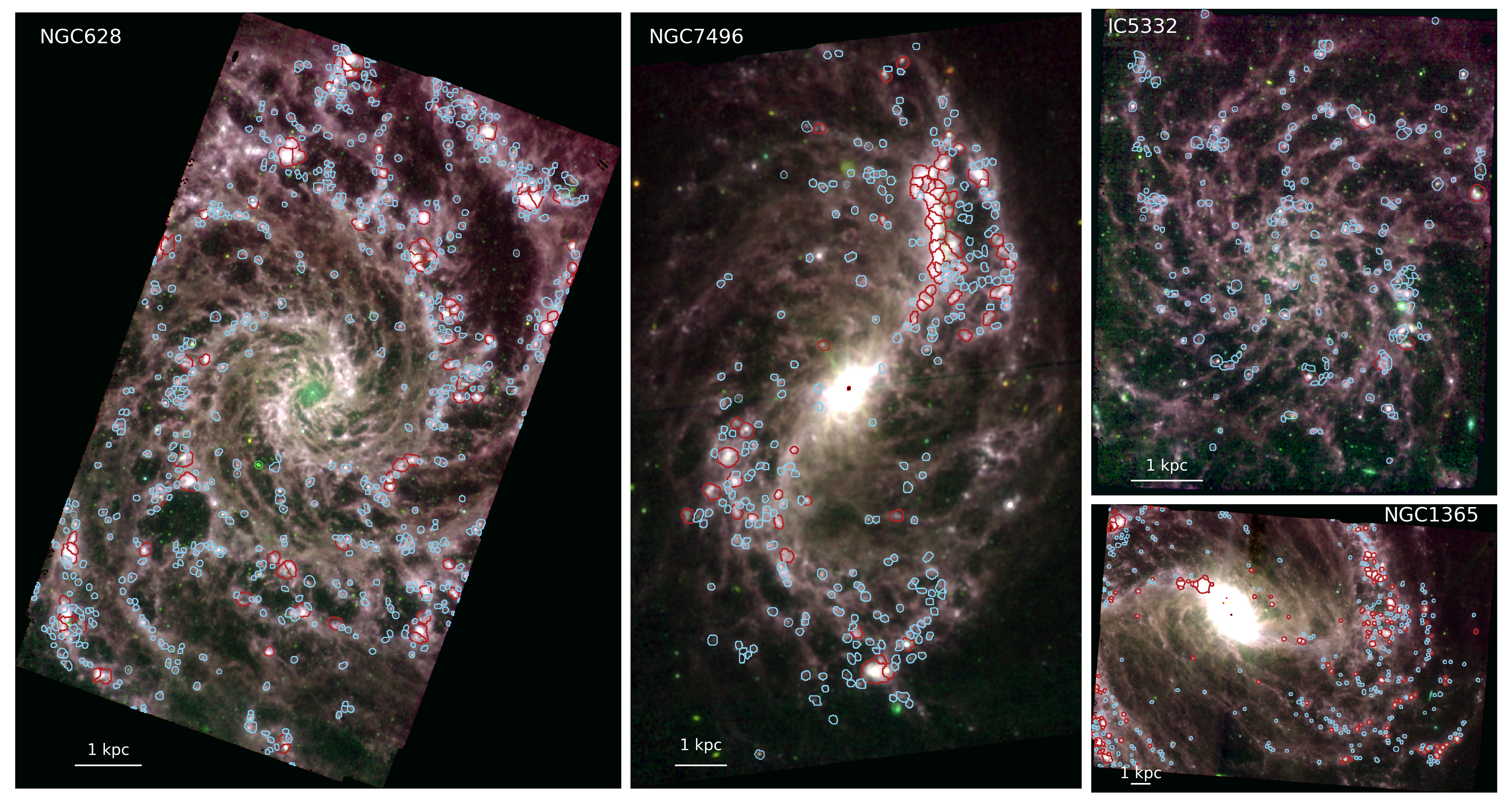}
    \caption{JWST/MIRI images in the F770W, F1000W and F1130W bands (red, green and blue channels, respectively) of the four galaxies analyzed in this letter. Contours show the borders of individual \HII\ regions selected from MUSE data, which meet the criteria in Sec.~\ref{sec:hii_selection}. Blue contours correspond to the \HII\ regions with hydrogen-ionizing photons rate $Q^0<10^{50}$ s$^{-1}$, and red contours to those with $Q^0>10^{50}$~s$^{-1}$. 
    }
    \label{fig:sample}
\end{figure*}

The properties of star-forming regions (e.g.\ age, metallicity, density) establish the balance between the PAH formation and destruction  processes. 
Multi-wavelength observations of \HII\ regions, their stellar populations and associated photo-dissociation regions (PDR) are therefore key for understanding the evolution of PAHs in the ISM of galaxies, and high ($<$100~pc) physical resolutions are critical to spatially isolating local from global effects \citep[see also][]{CHASTENET1_PHANGSJWST}. 
Until now, such observations were possible mainly in \HII\ regions in our Galaxy \cite[e.g.][]{Povich2007, Binder2018}, or very nearby galaxies \cite[e.g.][]{Bolatto2007, Sandstrom2010, Wiebe2011, Lebouteiller2011, Chastenet2017, Maragkoudakis2018, Chastenet2019, Mallory2022}, while most extragalactic studies 
were focused on integrated observations of entire galaxies or of large star-forming complexes \citep[e.g.][]{Engelbracht2005, Madden2006, Gordon2008, Khramtsova2013, Khramtsova2014, Maragkoudakis2018, Lin2020}. 

With the launch of the \JWST, it is now possible to investigate individual \HII\ regions and their immediate surroundings in significantly more distant galaxies than ever before, covering a more representative view of physical conditions in the ISM. Combining new \JWST\ data with the available high-resolution data in different wavelength ranges, we can probe now the evolution of the PAHs and other grains in connection with the star-forming  \HII\ regions. 

In this letter, we explore the abundance of PAH molecules within the boundaries of \HII\ regions at 28--64 pc resolution in four nearby star-forming galaxies  observed with \JWST\ as the first targets in the PHANGS-JWST treasury program \citep{LEE_PHANGSJWST}. We use MIRI images as tracers of emission from PAHs and very small dust grains. We combine these data with optical spectroscopy obtained from the MUSE integral-field spectrograph  \citep{Bacon2010} on the VLT as part of the PHANGS-MUSE program \citep{Emsellem2022}. Based on these data sets, we investigate the relation between the PAH fraction and properties of the \HII\ regions that could be potentially regulating their destruction (or formation). We refer the reader to the accompanying paper by \citet{CHASTENET1_PHANGSJWST}, where we analyze the PAH fraction on galaxy-wide scales.

This letter is organized as follows. In Section~\ref{sec:obs} we describe the underlying observational data. Section~\ref{sec:analysis} provides details on the methods and criteria implemented for selection of the \HII\ regions and deriving the properties of the gas, dust and young stars there. Section~\ref{sec:results} describes  results from the comparison of our tracer for PAH abundance and properties of the \HII\ regions. Section~\ref{sec:summary} summarizes our results.

\section{Observations}
\label{sec:obs}

\subsection{Galaxy sample}
\label{sec:sample}

We analyze \HII\ regions in the first four PHANGS-JWST galaxies. These galaxies have all been observed with VLT/MUSE, and thus are ideal objects to relate the properties of ionized gas to the PAH emission.
These nearby (D $\sim$ 9--20 Mpc) galaxies span an order of magnitude in stellar mass (see Table~\ref{tab:sample}), and almost two orders of magnitude in star formation rate (SFR). Two galaxies (NGC~1365 and NGC~7496) host active galactic nuclei (AGNs) and strong bars. NGC~628 is a prototypical grand-design spiral, while IC~5332 shows a flocculent spiral morphology. They represent a good cross-section of the PHANGS-JWST sample (Fig.~\ref{fig:sample}). 

\begin{table*}
\centering
\caption{Properties of the galaxies}

\label{tab:sample}
\begin{tabular}{cccccccccc} 
\hline
Galaxy & $D$ & $R_{25}^{\dagger}$ & $\log M_*^{+}$  & $\log$(SFR)$^{+}$   & 12+log(O/H)$^{*}$ & $\mathrm{N_{HII}}$\\ 
       & (Mpc) & (kpc) & ($M_\odot$) &  ($M_\odot\ \mathrm{yr}^{-1}$) & at $R_{\rm eff}$ &  \\
\hline 
IC5332 & ~9.01$^{1}$ & ~8.0 & 9.67 & $-$0.39  & 8.30 &  208 \\ 
NGC628 & ~9.84$^{1}$ & 14.1 & 10.34 & ~~0.24  & 8.48 &  668 \\ 
NGC1365 & 19.57$^{1}$ & 34.2 & 10.99 & ~~1.23  & 8.48 & 365  \\ 
NGC7496 & 18.72$^{2}$ & ~9.1 & 10.00 & ~~0.35 & 8.51 & 288 \\ 
\hline
\end{tabular}

\begin{footnotesize}
References: $^{\dagger}$\cite{Paturel2003}; $^{+}$\cite{Leroy2021}; $^{*}$Groves et al, submitted; \\ $^{1}$\cite{Anand2021a,Anand2021b}; $^{2}$\cite{Anand2021a,Shaya2017,Kourkchi2020}
\end{footnotesize}
\end{table*}

\subsection{PHANGS-JWST}\label{sec:obs:jwst}

The \JWST\ data were obtained in Cycle 1 as part of the PHANGS-JWST program (ID 02107; PI: J.~C.~Lee). The survey targets 19 nearby star-forming galaxies with NIRCam (F200W, F300M, F335M and F360M) and MIRI (F770W, F1000W, F1130W and F2100W) imaging. The first four galaxies have been observed, and their MIRI images (Fig.~\ref{fig:sample}) form the basis of our analysis. These galaxies have been imaged with two (for IC~5332 and NGC~7496), three (for NGC~628) or four (NGC~1365) MIRI pointings (88.8, 122.1, 310.8, 321.9 seconds per pointing in the F770W, F1000W, F1130W and F2100W bands, respectively, and with 4 dithers per pointing to properly sample the PSF), covering an area well-matched to the PHANGS-MUSE \citep{Emsellem2022} and PHANGS-HST observations \citep{Lee2022}. The sensitivity of the data is similar for all four galaxies in all MIRI bands ($0.11-0.15$~MJy~sr$^{-1}$ at $1\sigma$ level) except F2100W ($\sim0.25$~MJy~sr$^{-1}$).
We convolve all images to the PSF of the F2100W images (FWHM~$\simeq0.67''$; corresponding to 28--64 pc at the distance of our targets), as obtained via the \textsc{webbpsf} \citep{WebbPSF} model\footnote{https://webbpsf.readthedocs.io/en/stable/index.html}.  Convolution kernels were created following the procedure from \cite{Aniano2011}. 
Standard calibrations are applied, with minor modifications. A detailed description of the complete data reduction is presented in \citet{LEE_PHANGSJWST}.

\subsection{PHANGS-MUSE}\label{sec:obs:muse}
All the galaxies in the PHANGS-JWST program were previously observed with VLT/MUSE as part of the PHANGS-MUSE large program (PI: Schinnerer). The reduced data are publicly available\footnote{\url{https://archive.eso.org/scienceportal/home?data_collection=PHANGS}}, and the details of the observations and data reduction are given in \cite{Emsellem2022}. The angular resolution of the MUSE data for the objects in this study is $\sim 1''$ (45--95~pc), sufficient to isolate individual \HII\ regions from surrounding diffuse ionized gas (DIG). 

Based on MUSE H$\alpha$ emission line maps, a catalog of $\sim 30,000$ nebulae has been constructed (\citealt{Santoro2022}, Groves et al.\ submitted). The nebulae were selected using \textsc{HIIphot} \citep{Thilker2000}, which results in spatial masks that define the borders of each nebula. The catalog contains derived properties of each nebula (e.g., emission line fluxes, velocity dispersion, equivalent width, metallicity) based on their integrated spectra. The emission lines from each nebula are corrected for the effects of dust extinction using  the Balmer decrement, as described in Groves et al. In the remainder of this paper, we use the extinction-corrected values. \revone{We do not correct the measured fluxes for the contribution of the DIG. }
\HII\ regions are selected from the nebulae catalog based on the \cite{BPT} diagnostics considering \OIIIHb\ vs \NIIHa\ and \SIIHa\ line ratios, as described in Groves et al.\ (submitted, see also \citealt{Kewley2019} for a review). The present analysis is restricted to only those \HII\ regions  that meet stringent surface brightness and signal-to-noise (S/N) criteria (see Sec.~\ref{sec:hii_selection}). 

\section{Deriving the properties of the ionized gas and PAH\lowercase{s} in star-forming regions}
\label{sec:analysis}
\subsection{Selection of the nebulae and associated PAH emission}
\label{sec:hii_selection}

We use the spatial masks from the PHANGS-MUSE nebular catalog (see Sec.~\ref{sec:obs:muse}) to identify the positions of \HII\ regions in the \JWST\ data. To do that, we reproject these masks to the \JWST\ images.  
Since two of our galaxies contain an AGN, which produces hard ionizing radiation that can significantly affect the PAH fraction \citep[e.g.][]{Jensen2017, Lai2022, GB2022}, we exclude the centers of the galaxies from our analysis. 
Masks that isolate these particular environments are taken from \cite{Querejeta2021}. 

In order to minimize the DIG contribution to the regions' flux, which could bias derived properties of the nebulae, we select only those \HII\ regions with high H$\alpha$ equivalent width $EW(\mathrm{H\alpha})>16$\ \AA\ and H$\alpha$ surface brightness $\Sigma(\mathrm{H\alpha})>10^{39}\ \mathrm{erg\ s^{-1}\ kpc^{-2}}$
\citep[e.g.][]{Belfiore2022}. We also require $S/N > 5$ in every emission line considered in this letter, and $S/N > 15$ in the \Ha\ line. 
This ensures that we can recover accurate reddening-corrected fluxes of emission lines and  derived properties for each of the nebulae. 

In total, we select \nreg\ \HII\ regions across all four galaxies ($\mathrm{N_{HII}}$ in Tab.~\ref{tab:sample}) -- about 25\% of the \HII\ regions in our catalog 
(37\% for NGC~1365). We derive the ratio  $\log(\mathrm{\SIII9069,9532\AA/\SII6717,6731\AA})$\footnote{While \SIII$\lambda$9532 is outside the MUSE spectral range, the two \SIII\ lines show a fixed ratio, which we assume to be \SIII$\lambda9532\mathrm{\AA} \simeq 2.5\times$\SIII$\lambda$9069\AA \citep{Osterbrock2006}}, which is a proxy for the ionization parameter (the ratio of the local ionizing photon flux and the local hydrogen density; \citealt[][]{Kewley2002}), and we measure the oxygen abundance $12+\log\mathrm{(O/H)}$ (a proxy for gas-phase metallicity) with the S-calibration from \citet{Pilyugin2016}. We integrate the fluxes in each \JWST\ MIRI band within the apertures corresponding to the reprojected mask for each region. Some ratios of these mid-IR fluxes trace the PAH abundance in the ISM. We do not correct the measured fluxes for possible contribution from the diffuse ISM. \revone{Note, however, that removing the local background measured in circular apertures around \HII\ regions does not affect qualitatively the results presented below.}

\begin{figure*}
    \centering
    \includegraphics[width=\linewidth]{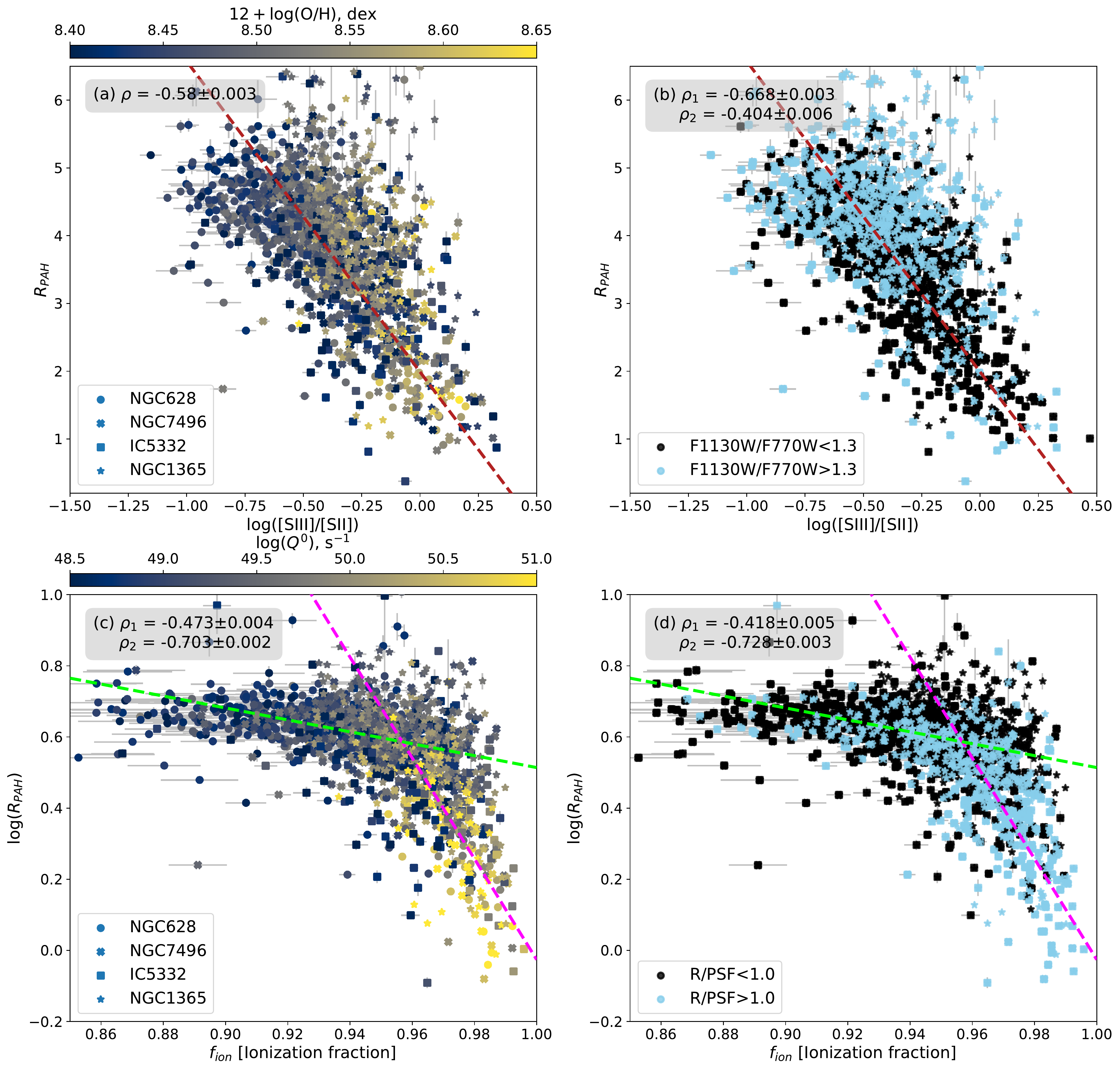}
    \caption{Correlation of PAH abundance, traced by $R_{\rm PAH}$, with  ionization parameter, traced by $\log$(\SIIISII) (Panels a, b), or ionization fraction $f_{\rm ion}$ (Panels c, d). Color encodes the oxygen abundance $12+\log(\mathrm{O/H})$ (Panel a), ratio of F1130W/F770W -- a proxy for the ionization state of PAHs (Panel b), number of hydrogen ionizing photons $Q^0$ (Panel c), and effective radii of the \HII\ regions relative to the PSF of the MUSE data (Panel d). Dashed lines show a linear fit for the \HII\ regions (Eq.~\ref{eq:corr1} for Panels a,b). Faint ($Q^0 < 10^{50}\ {\rm s}^{-1}$; green dashed line, Eq.~\ref{eq:corr_fnt}) and bright ($Q^0 > 10^{50}\ {\rm s}^{-1}$; magenta dashed line, Eq.~\ref{eq:corr_brt}) \HII\ regions are considered separately on Panels c and d. The Spearman's correlation coefficient ($\rho$) is given in the corner of each plot. \revone{Coefficients $\rho_1$ and $\rho_2$ are calculated separately for the black and cyan points (Panels b and d), or for the faint and bright regions (Panel c), respectively. The errors on $\rho$ are derived from 1000 Monte Carlo realization with the data randomly distributed around the measured values within the uncertainties.}}
    \label{fig:Rpah_vs_ip}
\end{figure*}

\subsection{Tracing the PAH fraction in the ISM}
\label{sec:pah_tracer}

The ratio of total fluxes in the \Spitzer\ bands $F_{8\mu m}/F_{24\mu m}$ is often considered as a tracer of \qPAH\ -- the mass fraction of PAH molecules with respect to the total dust mass  \citep[e.g.][]{Engelbracht2005, Bolatto2007, Sandstrom2010, Khramtsova2014, Oey2017}. \cite{Sandstrom2010} have shown that $F_{8\mu m}/F_{24\mu m}$ is indeed a good tracer of \qPAH, although it has a large scatter in the diffuse low-metallicity ISM where $q_{\rm PAH} < 1.5$\%. \cite{Khramtsova2013} found a stronger correlation between these parameters in star-forming regions.

In the case of \JWST, the flux of the {F2100W} band traces the emission from very small dust grains (similar to \Spitzer\ 24~$\mu$m) and correlates with the total IR luminosity ($F_\mathrm{TIR}$), whereas {F770W} and {F1130W} are dominated by PAH features at 7.7~$\mu$m (similar to \Spitzer\ 8~$\mu$m) and 11.3~$\mu$m, respectively, and trace the neutral and ionized ISM \citep{SANDSTROM1_PHANGSJWST}. 
Models by \cite{Draine2021} show that \qPAH\ is proportional to both $F_{7.7}/F_{\rm TIR}$ and $F_{11.2}/F_{\rm TIR}$. Using $F_{2100W}$ as a proxy for $F_{\rm TIR}$ introduces an unknown scaling, but provides a robust empirical tracer of \qPAH\ \citep{CHASTENET1_PHANGSJWST}. 
In this letter, we consider both the $7.7~\mu$m and $11.3~\mu$m PAH features and assume that:
\begin{equation}
\label{eq:rpah}
    q_{\rm PAH} \propto R_{\rm PAH} = \frac{F_{F770W}+F_{F1130W}}{F_{F2100W}}.
\end{equation}

\revone{Similarly to the works mentioned above, we assume further that $R_{\rm PAH}$ changes in \HII\ regions mostly due to variations of the PAH fraction regulated by the processes of their formation and destruction. Note, however, that redistribution of the dust grains of different sizes can also affect $R_{\rm PAH}$ if the relative abundance of very small grains is changing in \HII\ regions. For example, \citet{Everett2010} hypothesized that the very small grains could be re-supplied by the destruction of the dense embedded cloudlets overrun by the expansion of a bubble. Furthermore, radiation-pressure-driven shift or/and the stellar winds \citep{Gail1979, Draine2011} can redistribute the PAHs and very small grains, which has been suggested as a mechanism for evacuating the dust grains from the interior of several Galactic \HII\ regions \citep{Paladini2012}. \revtwo{Shell-like morphology of the Galactic \HII\ regions is indeed clearly seen in the mid-IR bands, especially in PAH-sensitive bands \citep[e.g.][]{Anderson2014}.} However, small grains ($<100$~\AA) drift much more slowly than larger grains, and the difference between the PAHs and the very small grains is insignificant \citep{Draine2011}.} \revtwo{Thus, these processes should not significantly change the relative fraction of PAHs within the \HII\ regions. Nevertheless, both of these effects can change the relative brightness of the rims of the \HII\ regions in the mid-IR bands. Given that we analyze the ratios of the fluxes integrated over the \HII\ regions, excluding the surrounding PDR (at least for the resolved \HII\ regions, see Sec.~\ref{sec:res:destruction}), our results presented below are unlikely to be affected by these processes.}     


\section{Results and discussion}
\label{sec:results}
\subsection{Destruction of PAHs by UV radiation}
\label{sec:res:destruction}

\begin{figure*}
    \centering
    \includegraphics[width=\linewidth]{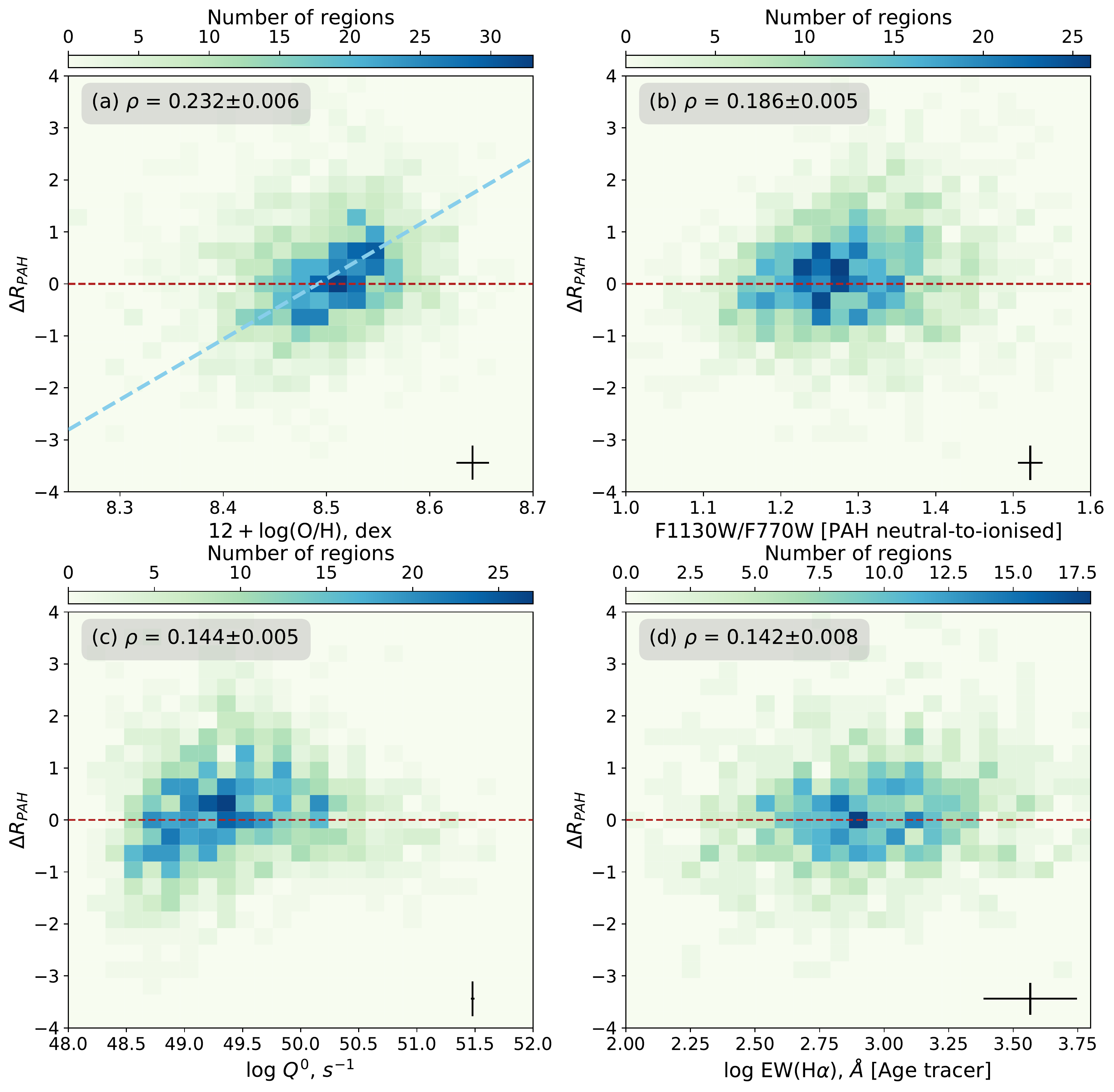}
    \caption{Secondary dependence of the PAH fraction on: the oxygen abundance of \HII\ regions, measured as $12+\log\mathrm{(O/H)}$ (Panel a); the F1130W/F770W flux ratio, sensitive to the relative fraction of neutral to ionized PAH (Panel b); the number of hydrogen ionizing photons $Q^0$ (Panel c); the equivalent width of \Ha, corrected for the contribution of the underlying old stellar population (Panel d). The 2D histograms on each panel show the distribution of  $\Delta R_{\rm PAH}$, obtained after subtracting of the values derived by Eq.~\ref{eq:corr1} from the measured $R_{\rm PAH}$. The red dashed line shows a zero-difference level. The light-blue dashed line on Panel (a) corresponds to a linear fit defined by Eq.~\ref{eq:corr_secondary}, tracing the secondary relation of $R_{\rm PAH}$ on metallicity.  The Spearman's correlation coefficient ($\rho$) is given in the corner of each plot. \revone{Its uncertainty is derived in the same way as in Fig.~\ref{fig:Rpah_vs_ip}. The error bars in the right-hand corner correspond to the 95th-percentile uncertainties of the measured values.}}
    \label{fig:Rpah_secondary}
\end{figure*}

PAH destruction is expected to be correlated with the hardness and intensity of the radiation field (see Sec.~\ref{sec:intro}). 
In previous works \citep[e.g.][]{Madden2006, Gordon2008, Lebouteiller2011, Maragkoudakis2018}, the flux ratio of the neon IR lines [Ne~\textsc{iii}]15.56~$\mu$m/[Ne~\textsc{ii}]12.81~$\mu$m was used as a tracer of the hardness of the ionizing radiation field. As shown in \cite{Kewley2019}, this ratio is dependent  on both the ionization parameter and the metallicity. In turn, the hardness of the radiation field depends on the metallicity \citep[e.g.][]{Groves2008}, and can be correlated with the ionization parameter in \HII\ regions \cite[see discussion in ][]{Kumari2021}. In this work, we use another tracer of the ionization parameter -- [S~\textsc{iii}]$\lambda$9069,9532\AA/[S~\textsc{ii}]$\lambda$6717,6731\AA, which can be measured in optical spectra. In contrast to [Ne~\textsc{iii}]/[Ne~\textsc{ii}], this ratio of sulfur lines is almost independent of  metallicity \citep{Kewley2019} and thus allows us to explore the relation of PAH abundances with the properties of the ionizing radiation field and the metallicity separately. 

In Fig.~\ref{fig:Rpah_vs_ip}(a, b) we demonstrate how $R_{\rm PAH}$ depends on $\log$(\SIIISII) in the \nreg\ \HII\ regions. The data shows a strong anti-correlation, 
implying that the PAH fraction in \HII\ regions is tightly related to the ionization parameter.  
This anti-correlation can be parameterized linearly as:
\begin{eqnarray}
R_{\rm PAH} = & (-4.58 \pm 0.16)\times \log(\mathrm{[S\ \textsc{iii}]/[S\ \textsc{ii}]}) + \label{eq:corr1} \\
& + (1.99 \pm 0.06) + \Delta R_\mathrm{O/H}, \nonumber  
\end{eqnarray}
where $\Delta R_\mathrm{O/H}$ accounts for any secondary dependence on metallicity. For now, we assume $\Delta R_\mathrm{O/H} = 0$ as we do not see a significant secondary correlation of $R_{\rm PAH}$ with $12+\mathrm{\log(O/H)}$ in Panel (a) (but see Sec.~\ref{sec:res:metallicity}). 

Panel (b) of Fig.~\ref{fig:Rpah_vs_ip} illustrates that the regions with higher $R_{\rm PAH}$ have, in general, a higher ratio of $F1130W/F770W \propto F_{11.3\mu m}/F_{7.7\mu m}$. This ratio is often used as a diagnostic of the neutral-to-ionized PAH ratio \citep[e.g.][]{Draine2001, Maragkoudakis2018}. Neutral PAHs have a higher ratio of $F_{11.3\mu m}/F_{7.7\mu m}$, which decreases as the fraction of ionized PAHs increases. Thus, the distribution of  $F1130W/F770W$ shown in this plot is consistent with a scenario where the PAHs are ionized and destroyed in the \HII\ regions with higher ionization parameter. Note, however, that this ratio can be affected by changes of the PAH size in higher intensity and harder radiation fields. The PAH feature at $3.3\mu m$, together with $F_{11.3\mu m}/F_{7.7\mu m}$, can help to disentangle the impact of PAH size from variations in the radiation field \citep{Draine2021, DALE_PHANGSJWST}. We refer the reader to \cite{CHASTENET2_PHANGSJWST}, who demonstrate that PAHs are not only more ionized, but also have smaller sizes in \HII\ regions, which is consistent with their gradual destruction in such environments.

Panel (c) of Fig.~\ref{fig:Rpah_vs_ip} provides some tentative conversions from the empirical correlations in Panels (a) and (b) towards a more physical interpretation. 
It shows $\log(R_{\rm PAH})$ vs $f_{\rm ion}$ -- the ionization fraction of hydrogen in each \HII\ region. The latter is calculated from $\log$(\SIIISII) using the empirical parametrization from \cite{Kreckel2022}. The color corresponds to the number of ionizing photons required to produce the observed \Ha\ luminosity $L(\mathrm{H}\alpha)$: $Q^0 \simeq 7.33\times10^{11}L(\mathrm{H}\alpha)$ assuming $T_e = 10 000$~K \citep{Osterbrock2006}, and with $L(\mathrm{H}\alpha)$ expressed in units of erg~s$^{-1}$. In this representation, two different sequences are suggested when examining the dependence of PAH fraction on $f_{\rm ion}$. Namely, $R_{\rm PAH}$ only slightly decreases with increasing $f_{\rm ion}$  (and ionization parameter) for fainter \HII\ regions, even when the ionization fraction is high. In turn, bright regions (likely associated with more massive star clusters) more strongly affect the PAH fraction. We perform separate linear fits to the faint and bright \HII\ regions (shown by green and magenta dashed lines in Fig.~\ref{fig:Rpah_vs_ip}c, respectively):
\begin{equation}
    \log(R_{\rm PAH})_{Q^0<10^{50}} = (-1.7 \pm 0.2) f_{\rm ion} + (2.2 \pm 0.2),
    \label{eq:corr_fnt}
\end{equation}
\begin{equation}
    \log(R_{\rm PAH})_{Q^0>10^{50}} = (-14.2 \pm 1.4) f_{\rm ion} + (14.2 \pm 1.3),
    \label{eq:corr_brt}
\end{equation}


Given the well-known size -- luminosity relation for \HII\ regions \citep[e.g.][]{Wisnioski2012}, a probable explanation of the bi-modality of the $\log(R_{\rm PAH})$ vs $f_{\rm ion}$ relation is a difference in their sizes. This difference can have both a physical or observational origin (or a combination). On the one hand, the larger, brighter regions might encompass the full scale height of the galactic disk and exhibit \qPAH~$\sim 0$ because the ISM is fully ionized towards these regions. On the other hand, the size of regions that are  \revtwo{unresolved} with MUSE might be overestimated, 
while for well-resolved regions the reprojected masks (see Section~\ref{sec:hii_selection}) should better isolate the \HII\ regions. \revtwo{Thus, in the unresolved regions one may expect to see a higher contribution from the diffuse ISM to the measurements. The effects leading to the redistribution of the dust grains at the outskirts of \HII\ regions (mentioned in Sec.~\ref{sec:pah_tracer}) may also be more prominent.} Panel (d) of Fig.~\ref{fig:Rpah_vs_ip} shows the same plot as Panel (c), but all points are color-coded by $R/PSF$ -- their effective angular radius relative to the resolution (FWHM of PSF) of the MUSE data. The color dashed lines show the regressions as on Panel (c) defined by Eqs.~\ref{eq:corr_fnt},\ref{eq:corr_brt}. Indeed, we start to resolve the \HII\ regions with MUSE at $Q^{0}\sim10^{50}$~${\rm s}^{-1}$, which naturally explains the bi-modality. \revone{Future \JWST\ observations of more nearby galaxies will help to clarify if the differences between the faint and bright \HII\ regions are real, or just an effect of the limited resolution.} Note that excluding \revtwo{the unresolved regions with} $R/PSF < 1$ (64\% of our sample) from consideration in Panels \revtwo{(a -- c) of Fig.~\ref{fig:Rpah_vs_ip}} does not affect our results, but the resolved regions do demonstrate a much tighter correlation with  $\log$(\SIIISII) \revtwo{(see Appendix~\ref{sec:app_subsamples})}.

\revtwo{In our measurements we do not attempt to correct for the contribution of the diffuse ISM to the optical emission line fluxes, or to the mid-IR bands, because performing a local background subtraction depends heavily on the exact \HII\ region boundaries we define and potentially introduces large uncertainties into our measurements. Nevertheless, we validate in Appendix~\ref{sec:app_subsamples} that our results do not change if we keep only those regions where the background emission is negligible compared to our measured fluxes. From this, we conclude that unresolved regions and contamination by the DIG/background mid-IR emission do not change our results, but are responsible for the scatter in Fig.~\ref{fig:Rpah_vs_ip}.}

\subsection{Dependence of the PAH abundance on metallicity}
\label{sec:res:metallicity}

The dependence of the PAH abundance on metallicity has been explored for galaxy-wide scales and for individual large star-forming complexes \cite[e.g.][]{Engelbracht2005, Khramtsova2013, Aniano2020}. While \qPAH\ is clearly systematically lower in dwarf galaxies compared to spiral galaxies, in the high-metallicity regime this relation exhibits a considerable scatter. The galaxies in our sample have relatively high metallicity, and cover only 0.3 dex of dynamic range (8.3 $<$ $12+\log\mathrm{(O/H)}$ $<$ 8.6), which could contribute to the lack of a clear secondary dependence of $R_{\rm PAH}$ with oxygen abundance in Fig.~\ref{fig:Rpah_vs_ip}a.

To explore the secondary dependence of $R_{\rm PAH}$ with metallicity, we subtract the relation in Eq.~\ref{eq:corr1} from our measurements and correlate the residuals $\Delta R_{\rm PAH}$ with 
$12+\log(\mathrm{O/H})$ in Fig.~\ref{fig:Rpah_secondary}a. We find a mild correlation between these values only considering the 2D histogram representing the statistical density of the points on this plot. This correlation can be parameterized linearly as:
\begin{eqnarray}
\Delta R_\mathrm{O/H} = & (11.3 \pm 0.5)\times ([12+\mathrm{\log(O/H)}] - 8.69) + \nonumber \\
& + (2.2 \pm 0.1). \label{eq:corr_secondary}  
\end{eqnarray}
Using this $\Delta R_\mathrm{O/H}$ in Eq.~\ref{eq:corr1}, we can describe the PAH fraction in \HII\ regions as a function of characteristics of the ionized gas -- its ionization parameter and metallicity. The dominant factor defining $R_{\rm PAH}$ is the ionization parameter, and the metallicity plays only a minor role leading to scatter in Fig.~\ref{fig:Rpah_vs_ip}a,b. Note that we do not find any prominent secondary relation of the PAH fraction on other parameters (F1130W/F770W, $Q^0$) explored in Fig.~\ref{fig:Rpah_vs_ip} (the bend in the trend for $Q^0$ in Fig.~\ref{fig:Rpah_secondary}c is probably due to the bi-modality discussed in Sec.~\ref{sec:res:destruction}).

\begin{figure*}
    \centering
    \includegraphics[width=\linewidth]{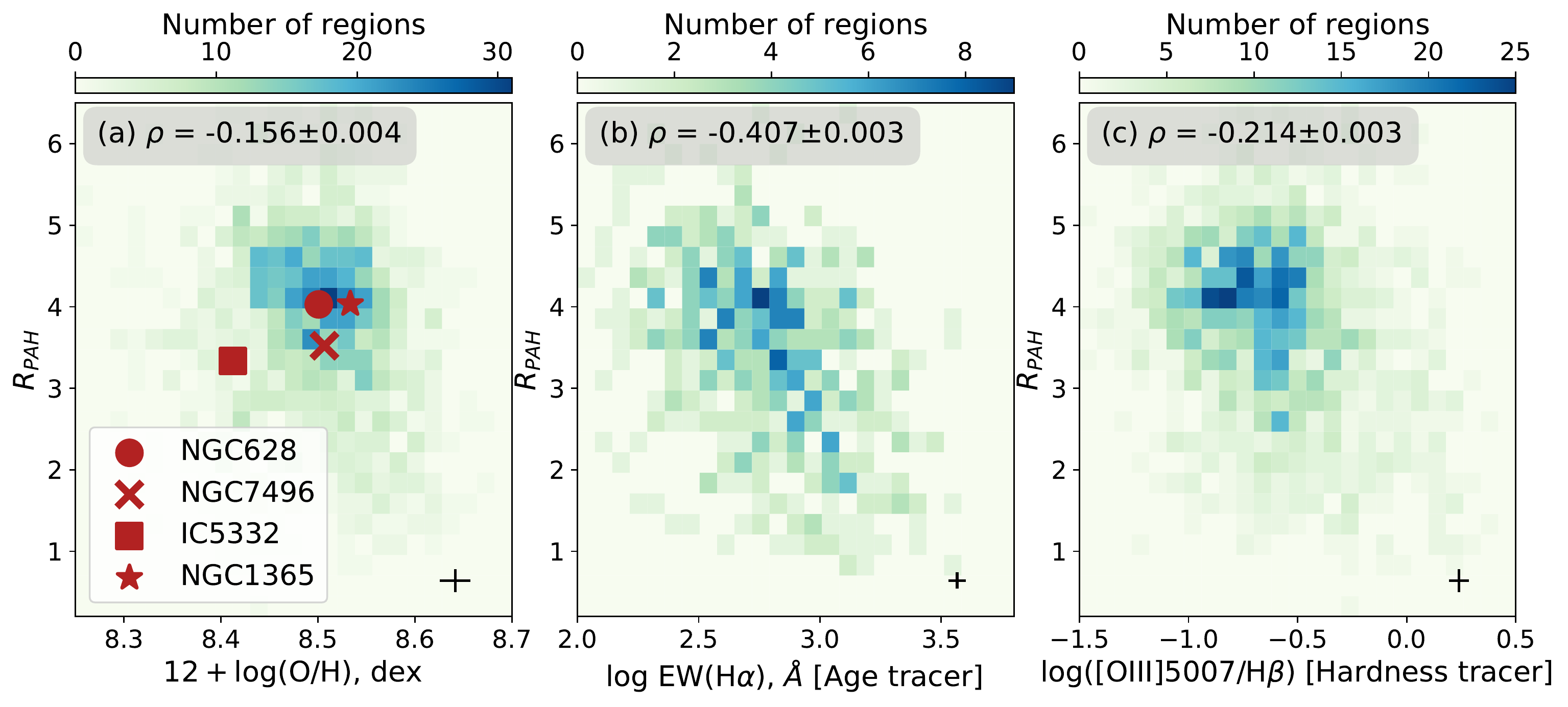}
    \caption{Correlation of the PAH fraction, traced by $R_{\rm PAH}$, with different physical conditions. Panel a: correlation with oxygen abundance, $12+\log(\mathrm{O/H})$, with red symbols showing the mean value for each galaxy. 
    Panel b: equivalent width of \Ha, $EW(\mathrm{H}\alpha)$, corrected for the contribution of the underlying old stellar population, and a tracer of \HII\ region age. Note that because of the large uncertainty introduced by this background continuum subtraction, we consider here only regions with relative error $\delta(EW(\mathrm{H}\alpha)) < 10\%$. 
    Panel c: $\log$(\OIIIHb), a tracer of the hardness of the UV radiation. \revone{The Spearman's correlation coefficient ($\rho$) is given in the corner of each plot. Its uncertainty is derived in the same way as in Fig.~\ref{fig:Rpah_vs_ip}. The error bars in the right-hand corner correspond to the 95th-percentile uncertainties of the measured values.}}
    \label{fig:Rpah_other_corr}
\end{figure*}

To demonstrate that we are not missing a primary correlation between $R_{\rm PAH}$ and metallicity, we show the 2D histogram distribution of these values in Fig.~\ref{fig:Rpah_other_corr}a. Red symbols on this plot correspond to the values averaged over all regions within each galaxy. These averaged values show an expected mild trend of increasing PAH fraction with the higher metallicity. At the same time, the distribution of the individual \HII\ regions demonstrates the opposite trend -- a very weak anti-correlation of $R_{\rm PAH}$ with oxygen abundance \citep[similarly to that for a small sample of high-metallicity galaxies in][]{Engelbracht2008}.

Overall, we conclude that the PAH destruction in \HII\ regions is nearly independent of metallicity, at least for the range of oxygen abundances we consider. Analyzing  highly-resolved data for the LMC and SMC, \cite{Chastenet2019} also found similar \qPAH\ in the luminous \HII\ regions of those two galaxies, despite their difference in metallicity and a difference in their global \qPAH\ values. This supports our claim that the weakness of the metallicity dependence is  not only due to the limited range of $12+\log(\mathrm{O/H})$ in our data. At the same time, \cite{CHASTENET1_PHANGSJWST} find a much stronger correlation between $R_{\rm PAH}$ and the metallicity of the diffuse ISM, where the intensity of \Ha\ is low. These results together imply that the metallicity regulates the PAH destruction and formation processes, but it becomes much less important in the \HII\ regions, where the presence of large amount of ionized gas and intense ionizing radiation plays a dominant role. 

\subsection{The role of UV hardness}
\label{sec:res:other_props}

The PAH fraction in \HII\ regions strongly correlates with the \SIIISII\ line ratio, with only a mild secondary dependence on the metallicity. While \SIIISII\ is a good tracer of the ionization parameter, it is also weakly dependent on the hardness of the UV radiation field (see Appendix~\ref{sec:app}). Many previous studies found a correlation between the PAH fraction and the hardness of the radiation field at the scale of star-forming complexes \citep[e.g.][]{Madden2006, Gordon2008, Khramtsova2014, Maragkoudakis2018}, and thus we cannot be sure that the trend seen in Fig.~\ref{fig:Rpah_vs_ip} is driven by the intensity of the hydrogen-ionizing radiation (or ionization parameter) and not by the hardness of the ionizing radiation field. 

We cannot directly measure the hardness of the UV field from our data, but as was shown by \cite{Levesque2010} and \cite{Khramtsova2014} (see also Fig.~\ref{fig:s3s2_vs_o3hb} in Appendix~\ref{sec:app}), \OIIIHb\ is a reasonable tracer of UV hardness at fixed metallicity. This ratio also depends on the ionization parameter, but to a lesser extent than \SIIISII. Given this, we would expect to see a strong correlation between $R_{\rm PAH}$ and \OIIIHb\ if the PAH destruction in \HII\ regions is regulated mostly by the hardness of the ionizing radiation field, and not by its intensity. In fact, we \revone{see only a relatively weak correlation} 
 between these parameters in Fig.~\ref{fig:Rpah_other_corr}c. Note that our result remains unchanged if we exclude IC~5332 from consideration (and only focus on the three galaxies of approximately similar metallicity). Thus, it is probable that it is indeed the ionization parameter, and not hardness, that drives the correlation in Fig.~\ref{fig:Rpah_vs_ip}.

\subsection{Towards an evolutionary sequence}
\label{sec:res:other_props2}
The dependence of $R_{\rm PAH}$ on the ionization parameter can be viewed within the context of an \HII\ region evolutionary sequence. 
We expect the ionization parameter to decrease with age, as older star clusters produce fewer ionizing photons and the \HII\ region expands \citep{Dopita2006}. The equivalent width of Balmer lines (e.g., $EW(\mathrm{H}\alpha$)) is another (more common)  indicator that decreases with  the age of an \HII\ region \citep{Levesque2013}. 
In Fig.~\ref{fig:Rpah_other_corr}b, we identify a correlation between $R_{\rm PAH}$ and $EW(\mathrm{H}\alpha$), where $EW(\mathrm{H}\alpha$) has been corrected for the contribution of the underlying old stellar population\footnote{This correction is performed by subtracting the corresponding continuum derived in a circular aperture surrounding each \HII\ region. For details, see Scheuermann et al. (\revone{submitted})}. $R_{\rm PAH}$ decreases with $EW(\mathrm{H}\alpha$), however the correlation is not as prominent as the correlation with \SIIISII. This suggests that the fraction of PAH molecules in the total dust mass increases slightly with time, if we assume that $EW(\mathrm{H}\alpha$) is a good tracer of the age of \HII\ region. Note that we do not find a secondary correlation between the residuals $\Delta R_{\rm PAH}$ (obtained after subtraction of the relation defined by Eq.~\ref{eq:corr1} from the observed data) on  $EW(\mathrm{H}\alpha$) (Fig.~\ref{fig:Rpah_secondary}d), probably because both $EW(\mathrm{H}\alpha$) and \SIIISII\ should decrease with the age. 
If such correlation with age is real, then it implies that the PAH fraction is higher in older regions. This can be explained if we assume that the balance between PAH formation and destruction changes with time. For example, the relative fraction of PAHs can increase with time if the very small grains have been destroyed or cleared from the older \HII\ regions more efficiently than PAHs, or if PAHs (even gradually) build up as result of destruction of larger dust grains (e.g., as results of shattering).

Previous studies have revealed a possible evolution of the PAH abundance with the age of star-forming regions, but the behavior is quite complex and depends on the balance between the processes of formation and destruction of the dust grains and PAHs \citep[e.g.][]{Wiebe2014, Khramtsova2014, Lin2020}. 
In particular, according to \cite{Khramtsova2014}, $R_{\rm PAH}$  increases with age for low-metallicity regions, and this trend flattens for higher metallicities. 
The correlation of $R_{\rm PAH}$ with $EW(\mathrm{H}\alpha$) in Fig.~\ref{fig:Rpah_other_corr}b is relatively weak, and $EW(\mathrm{H}\alpha$) itself is not always a strong tracer of \HII\ region age (Scheuermann et al., in prep.), thus we are hesitant to interpret this trend as strong evidence for an age sequence. Future work linking \HII\ regions with age dating of ionizing stellar clusters using PHANGS-HST and PHANGS-JWST data will provide a more robust quantification of these tentative age trends (see also \citealt{DALE_PHANGSJWST}).

\subsection{The role of shocks}

Beyond the correlations identified in this letter, we search for other properties regulating the PAH fraction on the scales of individual \HII\ regions. In particular, we tested whether shocks affect the PAH fraction by considering regions associated with high \Ha\ velocity dispersion (Egorov et al., in prep.) or classified as shock ionized by diagnostic optical emission lines ratios (Congiu et al., submitted), but did not find any significant differences from the trends we identify using the \HII\ regions in Fig.~\ref{fig:Rpah_vs_ip}. Thus, the ionization parameter (intensity of hydrogen-ionizing radiation) appears to be the dominant factor in defining the mechanisms for PAH destruction in \HII\ regions.

\section{Summary}
\label{sec:summary}

With \JWST\ it is now possible to systematically study the properties of the PAH component of the ISM at the scales of individual \HII\ regions in the galaxies beyond our Local Group, building a more representative picture of the interplay between ionizing sources and dust properties. As a pilot study, we analyze the PAH fraction in \nreg\ \HII\ regions within the disks of four nearby star-forming galaxies from the PHANGS-JWST program, and compare them with the properties of the ionized gas as obtained from PHANGS-MUSE observations. 
We find a strong anti-correlation between PAH fraction and ionization parameter, with a steeper dependence for more luminous regions. This is consistent with a scenario where the destruction of the PAH molecules is set by the hydrogen-ionizing UV radiation, though it is not clear if the observed trend reflects an evolutionary sequence. 
We find only a weak secondary dependence between PAH fraction and oxygen abundance, although we note that we cover a very limited range ($\sim$0.3 dex) in metallicity in our sample. Together with the results presented in \cite{CHASTENET1_PHANGSJWST}, this implies that, in contrast to the diffuse ISM, the metallicity becomes unimportant in defining the balance between PAHs formation and destruction in \HII\ regions, where the presence of ionized gas and hard radiation dominate. 

In this study leveraging new \JWST\ observations for four star-forming galaxies, we have demonstrated that hydrogen-ionizing UV radiation is the dominant mechanism for PAH destruction in \HII\ regions.


    
\section*{Acknowledgments}
This work is based on observations made with the NASA/ESA/CSA JWST. The data were obtained from the Mikulski Archive for Space Telescopes at the Space Telescope Science Institute, which is operated by the Association of Universities for Research in Astronomy, Inc., under NASA contract NAS 5-03127. The observations are associated with JWST program 2107. \revone{The specific observations analyzed can be accessed via \dataset[10.17909/9bdf-jn24]{http://dx.doi.org/10.17909/9bdf-jn24}.} 
Based on observations collected at the European Southern Observatory under ESO programmes 094.C-0623 (PI: Kreckel), 095.C-0473,  098.C-0484 (PI: Blanc), 1100.B-0651 (PHANGS-MUSE; PI: Schinnerer), as well as 094.B-0321 (MAGNUM; PI: Marconi), 099.B-0242, 0100.B-0116, 098.B-0551 (MAD; PI: Carollo) and 097.B-0640 (TIMER; PI: Gadotti). 

KK, OE and FS gratefully acknowledge funding from the Deutsche Forschungsgemeinschaft (DFG, German Research Foundation) in the form of an Emmy Noether Research Group (grant number KR4598/2-1, PI Kreckel). 
EJW, RSK, SCOG acknowledge funding from the Deutsche Forschungsgemeinschaft (DFG, German Research Foundation) -- Project-ID 138713538 -- SFB 881 (``The Milky Way System'', subprojects A1, B1, B2, B8, P1). 
HAP acknowledges support by the National Science and Technology Council of Taiwan under grant 110-2112-M-032-020-MY3.
TGW and ES acknowledge funding from the European Research Council (ERC) under the European Union’s Horizon 2020 research and innovation programme (grant agreement No. 694343).
MB acknowledges support from FONDECYT regular grant 1211000 and by the ANID BASAL project FB210003.
JMDK gratefully acknowledges funding from the European Research Council (ERC) under the European Union's Horizon 2020 research and innovation programme via the ERC Starting Grant MUSTANG (grant agreement number 714907).
COOL Research DAO is a Decentralized Autonomous Organization supporting research in astrophysics aimed at uncovering our cosmic origins.
E.C. acknowledges support from ANID Basal projects ACE210002 and FB210003.
FB would like to acknowledge funding from the European Research Council (ERC) under the European Union’s Horizon 2020 research and innovation programme (grant agreement No.726384/Empire)
MC gratefully acknowledges funding from the DFG through an Emmy Noether Research Group (grant number CH2137/1-1).
RSK and SCOG acknowledge support from the European Research Council via the ERC Synergy Grant ``ECOGAL'' (project ID 855130), from the Heidelberg Cluster of Excellence (EXC 2181 - 390900948) ``STRUCTURES'', funded by the German Excellence Strategy, and from the German Ministry for Economic Affairs and Climate Action for funding in project ``MAINN'' (funding ID 50OO2206).
ER acknowledges the support of the Natural Sciences and Engineering Research Council of Canada (NSERC), funding reference number RGPIN-2022-03499.
KG is supported by the Australian Research Council through the Discovery Early Career Researcher Award (DECRA) Fellowship DE220100766 funded by the Australian Government. 
KG is supported by the Australian Research Council Centre of Excellence for All Sky Astrophysics in 3 Dimensions (ASTRO~3D), through project number CE170100013. 
G.A.B. acknowledges the support from ANID Basal project FB210003.
JC acknowledges support from ERC starting grant \#851622 DustOrigin.
AKL gratefully acknowledges support by grants 1653300 and 2205628 from the National Science Foundation, by award JWST-GO-02107.009-A, and by a Humboldt Research Award from the Alexander von Humboldt Foundation.
JPe acknowledges support by the DAOISM grant ANR-21-CE31-0010 and by the Programme National ``Physique et Chimie du Milieu Interstellaire'' (PCMI) of CNRS/INSU with INC/INP, co-funded by CEA and CNES.

\facilities{JWST(MIRI), VLT(MUSE)}

\software{Astropy \citep{2013A&A...558A..33A,2018AJ....156..123A}, Cloudy \citep{Ferland2017}, pyCloudy \citep{Morisset2013}, Starburst99 \citep{Leitherer1999, Leitherer2014}, WebbPSF \citep{WebbPSF}}







\appendix
\section{Physical interpretation of emission line ratios in the context of photoionization modeling} 
\label{sec:app}

\begin{figure*}
    \centering
    \includegraphics[width=\linewidth]{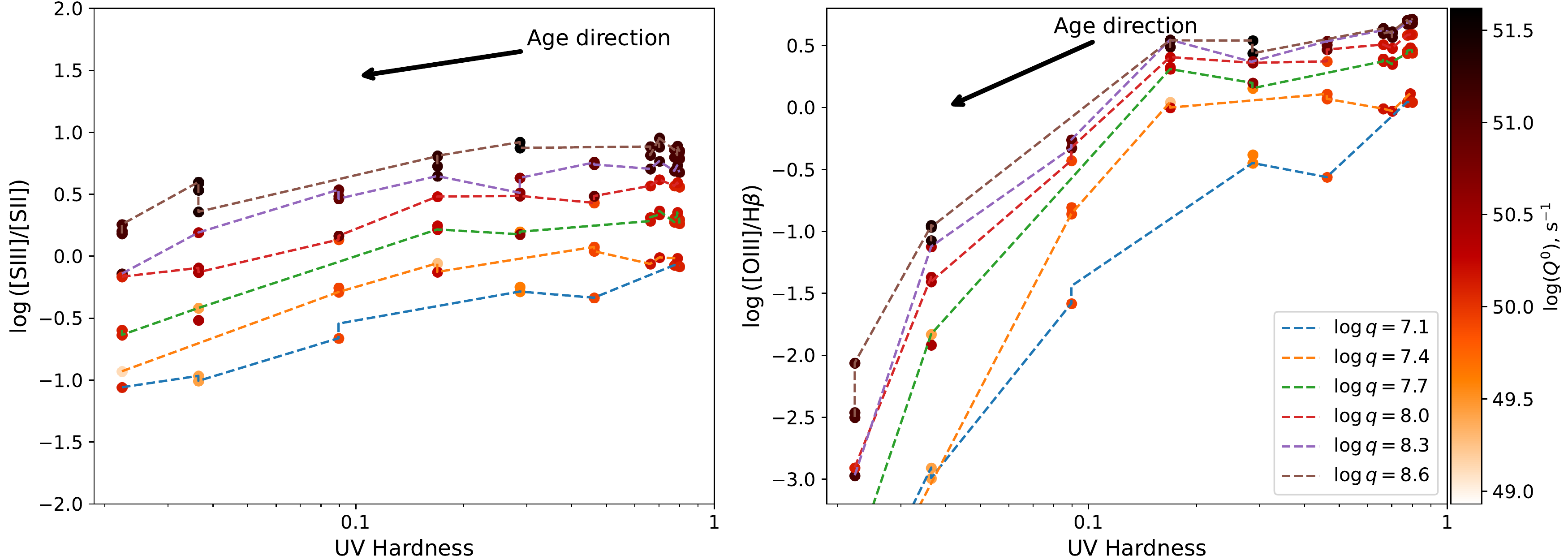}
    \caption{Dependence of \SIIISII\ (left-hand panel) and \OIIIHb\ (right-hand panel) optical emission line ratios on the hardness of the ionizing UV radiation and on the ionization parameter $q$, according to a grid of \textsc{starburst99} and \textsc{cloudy} models (see text). Different points are the individual nodes of the model grid, defined by the different mass and age of the ionizing cluster and the density of the \HII\ region. The color of the points encodes the number of hydrogen-ionizing photons,~$Q^0$. }
    \label{fig:s3s2_vs_o3hb}
\end{figure*}

Here we consider how the hardness and intensity of the ionizing radiation impact the \SIIISII\ and \OIIIHb\ line ratios, and how they evolve with the age of the \HII\ regions. \cite{Kreckel2022} presented \textsc{starburst99} \citep{Leitherer1999, Leitherer2014} and \textsc{cloudy} \citep{Ferland2017} models computed for a grid of star clusters and \HII\ region physical properties matched to the observed properties as derived in the PHANGS-MUSE and PHANGS-HST data. Here we use a sub-set of their models for metallicity $Z\sim0.6Z_\odot$ (approximately equal to the metallicity of the four galaxies considered in this work, see Table~\ref{tab:sample}). We consider  clusters of mass $\log(M_*/M_\odot) = 3.5, 4.5, 5.5$ and age  $t = 1...9$~Myr (with step of 1~Myr) as the ionizing source of a cloud with constant hydrogen density $n_{\rm H} \simeq n_{\rm e} = 20, 90, 300$~cm$^{-3}$. For simplicity, here we consider only one nebular geometry -- a shell with an inner radius of 15~pc, and filling factor $ff=1$. 

Similar to \citet{Khramtsova2014}, we parameterize the hardness of the UV radiation as the ratio of the total flux in the ionizing continuum at short wavelengths (20~\AA~to 912~\AA) to that at longer wavelengths (912~\AA~to 2000~\AA). Fig.~\ref{fig:s3s2_vs_o3hb} shows how both ionization parameter and UV hardness are traced by the emission line ratios \SIIISII\ and \OIIIHb. These line ratios are sensitive to both parameters, but \SIIISII\ shows a stronger dependence on the ionization parameter, while \OIIIHb\ is less sensitive to ionization parameter but varies significantly with the hardness of the UV radiation. Thus, by analyzing the dependence of PAH fraction in \HII\ regions using both of these line ratios, we can  disentangle whether UV hardness or the number of ionizing photons (proportional to the ionization parameter) is the main driver for PAH destruction. We consider this in Sec.~\ref{sec:res:other_props}.
 
\section{\revtwo{Testing the impact of limited angular resolution and contamination by emission from the diffuse ISM}}
\label{sec:app_subsamples}

\revtwo{As we show in Section~\ref{sec:res:destruction}, a large fraction of the \HII\ regions analyzed in this paper are not well-resolved by our PHANGS-MUSE observations ($\rm R/PSF < 1$ in Fig.~\ref{fig:Rpah_vs_ip}d). Also, we do not correct the measured fluxes of the optical emission lines and mid-IR bands for a local background, as implementing such subtraction is a fairly uncertain process. Nevertheless, both these factors may affect our measurements. In order to check that our conclusions are not biased because of these factors, we perform several tests in this Section. }

\begin{figure*}
    \centering
    \includegraphics[width=\linewidth]{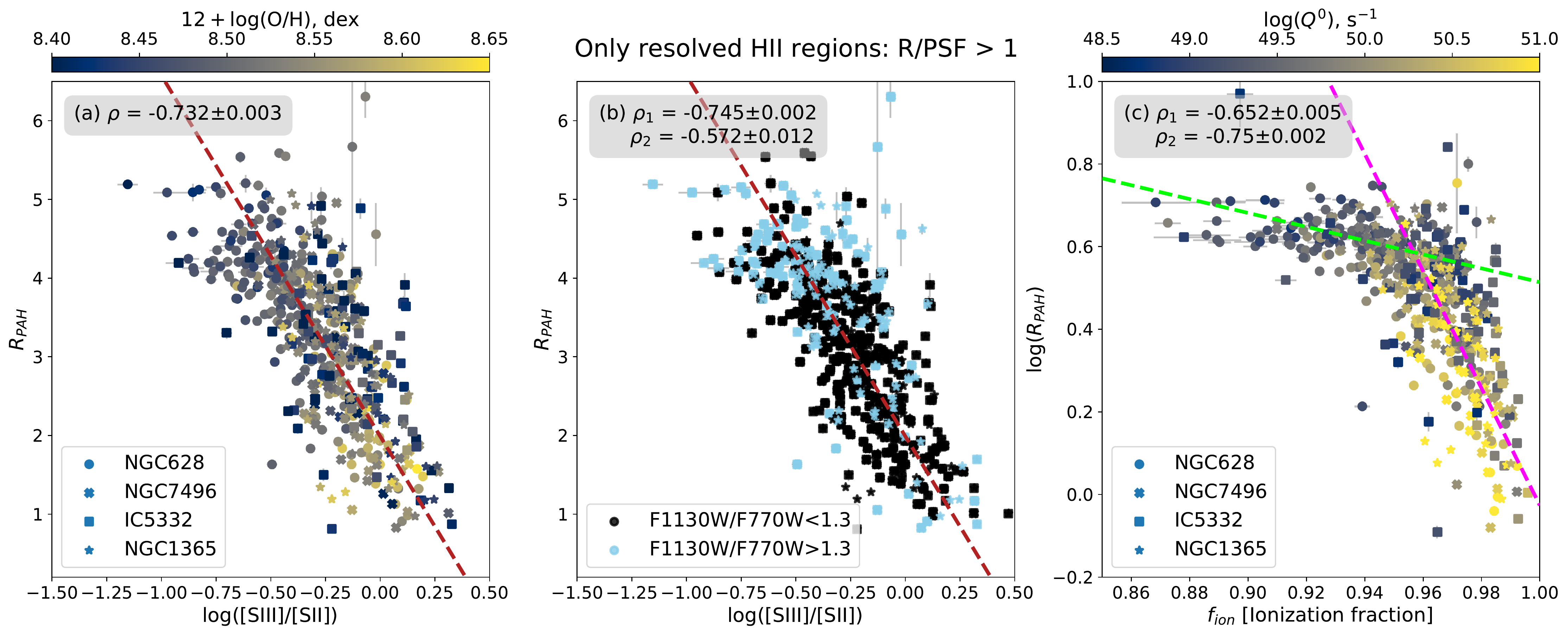}
    \caption{\revtwo{Same as in Fig.~\ref{fig:Rpah_vs_ip}(a--c), but for the resolved \HII\ regions only (those with $\rm R/PSF > 1$ in Fig.~\ref{fig:Rpah_vs_ip}d). The colored lines show the same linear regressions as in Fig.~\ref{fig:Rpah_vs_ip}.}}
    \label{fig:check_size}
\end{figure*}

\revtwo{In unresolved \HII\ regions, the measured value of $\rm R_{PAH}$ may be biased because of a larger contribution from the surrounding PDR and diffuse ISM to the aperture, which is indeed seen in Fig.~\ref{fig:Rpah_vs_ip}d. In Fig.~\ref{fig:check_size} we repeat the analysis presented in  Sec.~\ref{sec:res:destruction}, but consider only the resolved \HII\ regions having $\rm R/PSF > 1$ (44\% of total number of \HII\ regions in the analysis). As shown, the relations drawn from the entire sample are the same (but tighter) as the relations observed for the resolved \HII\ regions alone.}

\revtwo{The second test shown in Fig.~\ref{fig:check_size} allows us to confirm that the contamination by the diffuse ISM does not introduce a bias to our results. Here we follow the same approach as in \cite{Kreckel2022} -- we consider only the \HII\ regions with a significant contrast in the \Ha\ and \SII\ fluxes and in the mid-IR bands compared to the local background. We estimate the contrast as the maximal value of $C = F_{HII}/(I_{bgr} \times S)$ among all considered lines/bands, where $F_{HII}$ is a measured integral flux of the \HII\ region, $I_{bgr}$ is the median brightness of the surrounding local background estimated across a 10$\arcsec$ circular aperture after masking all \HII\ regions, and $S$ is the area within the \HII\ region mask. In Fig.~\ref{fig:check_size} we show only regions with a contrast $C > 2$, for which the contribution of the diffuse emission to the total flux is negligible ($\sim$29\% of the entire sample). This `clear' sample demonstrates the same trends as the entire sample in Fig.~\ref{fig:Rpah_vs_ip}.}

\revtwo{We may thus conclude that the presence of unresolved \HII\ regions and contamination by the diffuse ISM (which is still unavoidable for galaxies outside the Local Group) does not affect the results of our analysis, but add scatter to the observed relations. Note, however, that when considering these sub-samples we cannot judge how reliable our findings are regarding the secondary dependence of the PAH fraction on metallicity (Sec.~\ref{sec:res:metallicity}, Fig.~\ref{fig:Rpah_secondary}) because the number of regions at the low-metallicity end of the `clear' samples analyzed is too small.}

\begin{figure*}
    \centering
    \includegraphics[width=\linewidth]{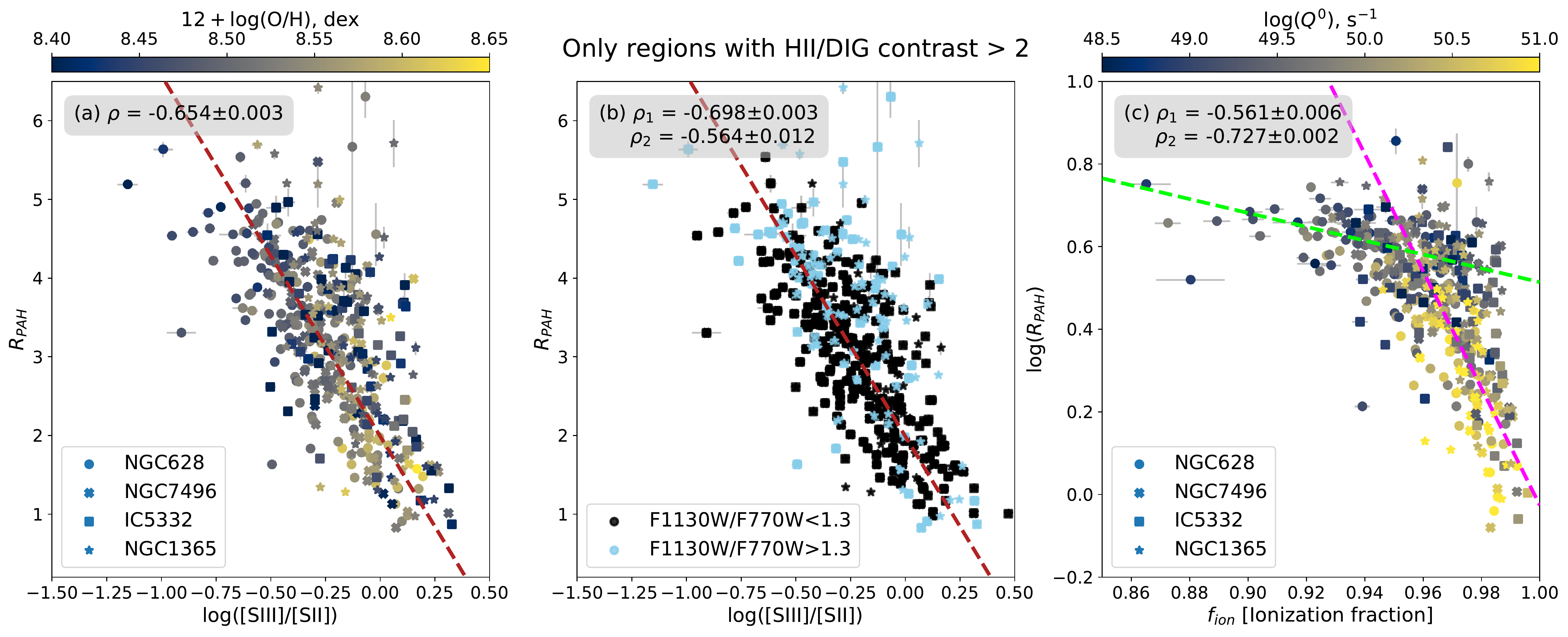}
    \caption{\revtwo{Same as in Fig.~\ref{fig:check_size}, but for a sub-sample of the \HII\ regions with a large contrast against the surrounding local background in the H$\alpha$, \SII\ emission lines and mid-IR bands.}}
    \label{fig:check_dig}
\end{figure*}

\bibliography{PAH_references, phangsjwst}{}
\bibliographystyle{aasjournal}



\suppressAffiliationsfalse
\allauthors

\end{document}